# Maintenance Duration Estimate for a DEMO Fusion Power Plant, based on the EFDA WP12 pre-conceptual studies

O. Crofts[a], J. Harman[b].

[a]EURATOM/Culham Center Fusion Energy, CulhamScience Centre OX14 3DB Abingdon, UK
[b]EFDA Close Support Unit Garching, Boltzmannstaße 2, D-85748, Garching Bei München, Germany

The erosion and high neutron flux in a fusion power plant results in the need for frequent remote replacement of the plasma facing components. This is a complex and time consuming remote handling operation and its duration directly affects the availability and therefore the commercial viability of the power plant.

A tool is needed to allow the maintenance duration to be determined so that developments in component design can be assessed in terms of their effect on the maintenance duration. This allows the correct balance to be drawn between component cost and performance on the one hand and the remote handling cost and plant availability on the other.

The work to develop this tool has begun with an estimate of the maintenance duration for a fusion power plant based on the EFDA DEMO WP12 pre-conceptual design studies [1]. The estimate can be readily adjusted for changes to the remote maintenance process resulting from design changes. The estimate uses data extrapolated from recorded times and operational experience from remote maintenance activities on the JET tokamak and other nuclear facilities.

The Power Plant Conceptual Study from 2005 [2] proposes that commercial viability of a power plant would require an availability of 75% or above. Results from the maintenance estimate described in this paper suggest that this level of availability could be achieved for the planned maintenance using a highly developed and tested remote maintenance system, with a large element of parallel working and challenging but feasible operation times.

## 1. Introduction

The plasma facing components in a fusion power plant are exposed to a very high neutron flux. This results in progressive changes to the material properties. Ultimately the material properties of the plasma facing materials are sufficiently degraded to require the replacement of the Multi-Module Segment (MMS) breeding blankets and the divertor cassettes.

The amount of time that a fusion power plant can be generating electricity and therefore revenue is dependent upon the availability of the plant [2]. A minimum availability is therefore required for commercial viability.

The main factor affecting the availability of a fusion power plant will be the duration of the periodic scheduled remote maintenance required to replace the plasma facing components during the life of the plant. This is because the remote maintenance is complex and time consuming and it is a significant challenge to be overcome in the realisation of fusion power.

It is therefore vital to have an understanding of the duration of the scheduled remote maintenance and to have a tool to allow the impact of component designs to be assessed in terms of their effect on the maintenance duration.

The maintenance duration assessment tool will allow the duration of a range of component design options to be assessed. This allows the correct balance to be drawn between component performance and maintenance duration to maximise the commercial viability of the power plant.

The work to develop this tool has begun with the development of a bottom-up estimate of the maintenance duration for the EFDA DEMO 2012 fusion power plant [1]. The estimate can be readily adjusted for changes to the remote maintenance process resulting from design changes.

This paper describes the development of the estimate in which the duration of individual operations are summed to provide the total duration for a range of maintenance scenarios. It takes into account factors such as operator efficiency and the reliability and recovery of remote handling systems and tools.

## 2. Remote maintenance

The scheduled remote maintenance operations considered in this estimate are for the vertical maintenance system developed for the EFDA WP12 pre-conceptual studies [1, 3].

In this system, all sixteen upper ports are used for the removal of the MMS blankets and all sixteen divertor ports are used for the removal of the divertor cassettes.

---

*author's email: oliver.crofts@ccfe.ac.uk*

This is required to minimise the in-vessel remote handling operations and to maximise the amount of parallel working that can be undertaken.

A range of casks are docked to the bio-shield and vessel ports at the upper port and at the divertor port to deploy remote handling systems and to deliver and remove components. During remote maintenance operations, the Active Maintenance Facility [4] supplies and receives casks containing remote handling equipment and components from a buffer store.

A section through the proposed vertical maintenance system is shown in figure 1.

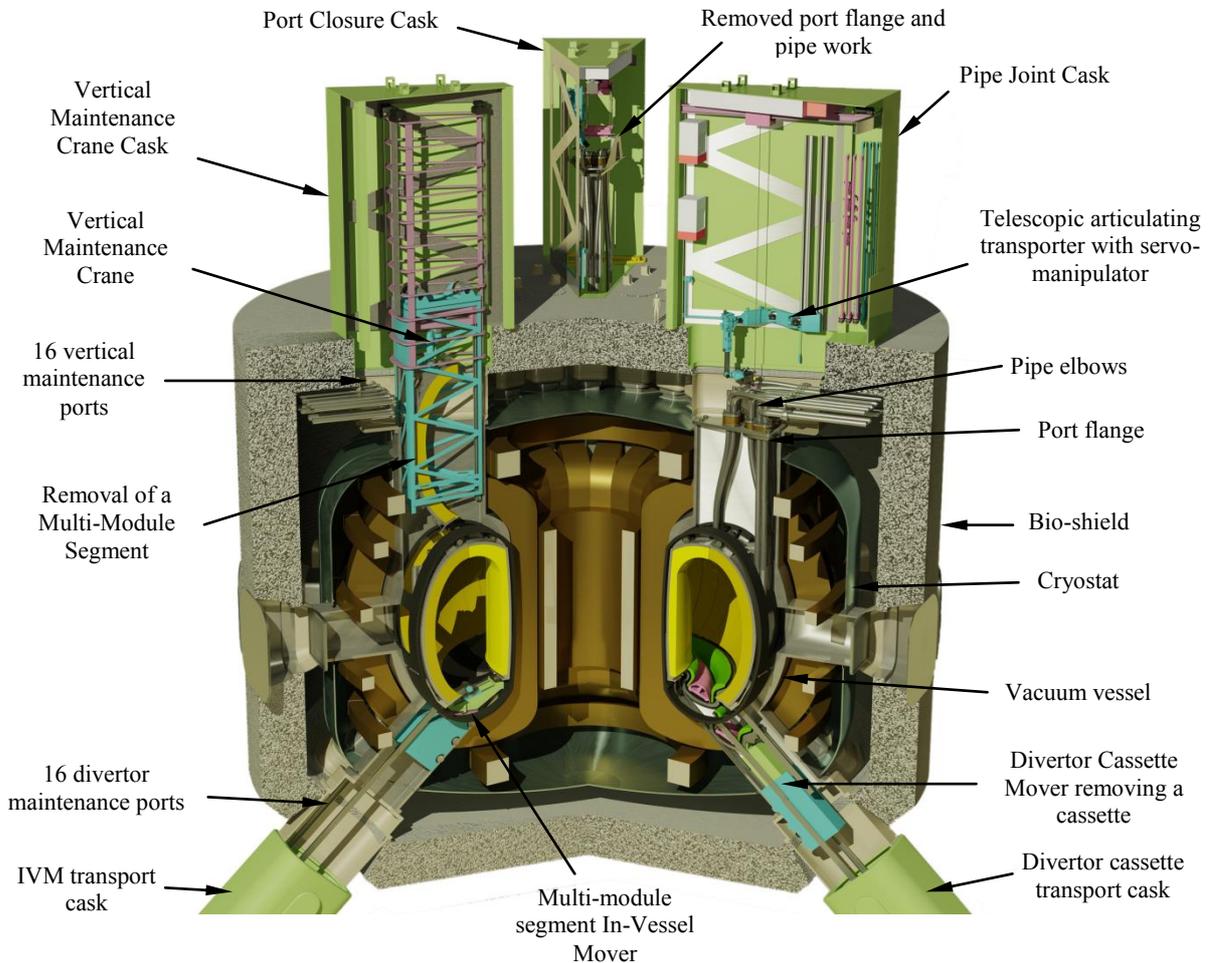

Figure 1: Section through the proposed DEMO Vertical Maintenance System

**2.1 Sequence of operations**

The divertor cassettes must be removed to allow the in-vessel mover to be deployed during MMS blanket removal, but since the blanket life is likely to be higher than the divertor life this will not add to the number of times the divertor is removed.

The sequence of operations at any port is to prepare the port, replace the plasma facing components and then seal the port. This is described in detail in the EFDA DEMO WP12 pre-conceptual design studies [1] and is summarised below.

Preparing a port involves removing the port shield plug, deploying the pipe joint cask and then deploying the port closure cask. The pipe joint cask remote handling system removes the elbows in the pipe connections to allow access to the port flange and cuts the pipe connections close to the component. The port closure cask then removes the port flange with pipes attached allowing access to the components below. Sealing the port is the reverse operation.

Replacement of the MMS blanket segments involves deploying the vertical maintenance crane from a cask above the blankets and the in-vessel mover from a divertor cask. The in-vessel mover disconnects the blanket from the vessel and assists in moving the blanket away from the shear key locations. The vertical maintenance crane then extracts the blanket into the cask for removal to the Active Maintenance Facility. Once all five blankets from the sector are removed they are replaced with new components in the reverse operation.

The operations between the vertical port and the divertor port in one sector are tied together by the replacement of the MMS blanket segments because of the requirement to deploy the in-vessel mover. See figure 2 below.

The replacement of the divertor cassettes involves the deployment of three divertor transport casks from which a cassette mover is deployed and a cassette is extracted and transferred to the Active Maintenance Facility. Once all three cassettes have been removed and the blankets exchanged if required, three new cassettes are installed in the reverse operation.

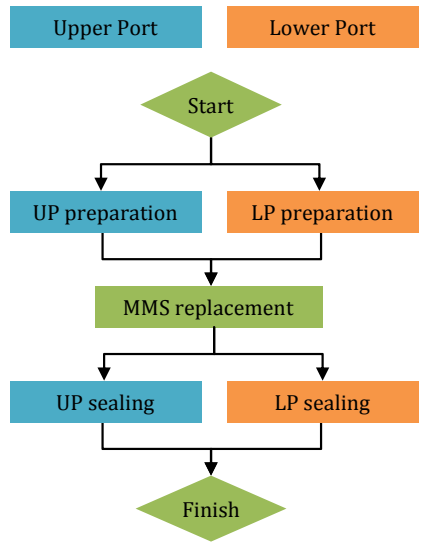

Figure 2: Sequence flow chart for the maintenance operations between the upper and lower ports

**2.2 Maintenance scenarios**

A number of different maintenance scenarios were considered, namely the unplanned replacement of a single blanket or cassette due to failure and the planned maintenance of the divertor or the divertor and the blankets.

The remote operations at each sector are essentially independent. Therefore by increasing the number of remote handling systems it is possible to operate on multiple sectors in parallel. The estimate considered the maintenance duration for 1, 2, 4, 6 and 8 sectors being maintained in parallel.

# 3. Calculation of the maintenance duration

The operation durations for each of the cask systems were calculated using a bottom-up approach, where durations are estimated for each sub-task and these were summed to produce the total operation duration.

This work is described in detail in the Operational Efficiency Design Assessment Study for 2012 [5] and is summarised below.

The duration for individual operations such as welding or bolting therefore occur in only one place in the calculation and can be adjusted with updated data or for sensitivity analysis.

The operation durations for each cask system were then combined to provide the total duration for the port opening and closing and for the component replacement. Finally these durations were combined in different ways to produce the overall maintenance duration for the different maintenance scenarios, such as for the removal of a single plasma facing component or for the parallel operation on multiple sectors at once.

**3.1 Input Data**

The duration of operations is based on and extrapolated from recorded times and from operational experience at JET and other nuclear facilities [6, 7] and have been reviewed by experienced operations staff at JET.

**3.2 Recovery duration**

Due to the failure rate of complex remote handling equipment and the duration to affect recovery operations the maintenance duration estimate would be overly optimistic if an estimate for the recovery duration was not included.

To estimate the recovery duration, first the additional tasks required to recover from the range of failure scenarios were considered, then the probability of the events that could lead to the scenarios were summed because the events are independent.

The recovery scenarios considered were replacing a cask containing failed equipment, including exchange of the cask component inventory if necessary, and the replacement of pipe groups containing a failed weld.

Only significant failures on the critical path were considered that would result in the replacement of a cask system or a component. It was assumed that most components prone to failure will have redundancy built in and that spare tools will be available in the casks.

It was also assumed that the recovery operations follow previously determined, well developed procedures. Highly unlikely failures that would require special recovery procedures for investment protection were not included.

When considering multiple remote handling systems working in parallel, the recovery duration is the time it takes to complete the longest duration recovery for any one of the systems working in parallel. This extended duration was calculated by combining the distributions from the separate events to find the typical longest delay from any of the separate systems.

**3.3 Cask transporter utilisation**

The upper port casks are transported by a crane and the divertor port casks by a floor mounted transporter. There is only one crane and one transporter for all cask moves so on occasion the cask transport system will not be available when required. This will add to the transporter operation duration.

The average transporter operation duration was estimated using the stochastic M/D/1 queue theory formula shown below for a single server with fixed (deterministic) job service time D, where µ is the serving rate (1/D) and ρ is the transporter utilization fraction.

$$\frac{1}{\mu} \cdot \frac{2-\rho}{2-2\rho}.$$

Delays due to the cask transporters are low at less than 1% of the total maintenance duration. This is primarily due to the low job service time of 30 minutes. Transporter utilisation is around 10% for a single remote handling system, rising to just over 80% for 8 systems operating in parallel.

### 3.4 Efficiency factors

The operation durations were multiplied by two factors to turn the ideal durations into the real durations observed in practice during remote handling operations.

The first efficiency factor was a shift productivity ratio. Various shift patterns were considered. A two shift operation pattern was adopted as the most likely option because it provides 16 hours operation per day compared to only 8 for a single shift. It is also easier to implement and more efficient in terms of the shift teams required using 3 shift teams rather than 5. The least productive hours in the middle of the night are also not used for remote operations but can be used for housekeeping operations as required.

The second factor was the operator productivity factor. No team of operators can provide 100% productivity. Reduced productivity occurs due to a large number of factors, the most significant being human error which is highest for man-in-the-loop remote handling operations, particularly at the beginning of operations or when a high number of repetitions are being completed.

At JET, productivity varies considerably depending on the task, but a figure of about 70% is considered suitable for well-developed procedures with man-in-the-loop control. These procedures include the tool deployment operations considered for DEMO [7].

It is estimated that the automatic procedures would have a productivity of about 90%. These procedures include removal and replacement of the MMS blanket modules and divertor cassettes and the docking of casks.

### 3.5 Assumptions

All sixteen upper ports and all sixteen divertor ports are available for remote maintenance operations.

Operation durations assume a highly developed and tested remote handling system with high productivity and reliability, operated by skilled and experienced staff.

All recovery operations follow well developed and tested procedures.

The Active Maintenance Facility is capable of supplying and receiving casks without delaying the critical path remote handling operations and has a suitable number of spares to account for all likely failure scenarios.

Redundancy is applied to drives and systems and spares tools are provided where ever possible.

Health physics checks are conducted by automated instrumentation so approval to continue operations can be given using readouts at the operator's control station.

All remote handling control systems are pre-commissioned in the Active Maintenance Facility and casks have shielded local control systems that remain powered-up so that the cask plug connections when connected to the bio-shield and vacuum vessel are relatively small and simple. It also minimises the duration of pre-operation checks and commissioning.

TIG welding is used as the pipe joining technology due to its well characterised performance with a triple weld head employed on large bore pipes to reduce weld duration.

## 4. Results

The results from the estimate are presented in detail in the final report for the Operational Efficiency Design Assessment Study for 2012 [5]. Result highlights are given in table 1 and in the paragraphs below.

The time taken to replace the blankets and divertor cassettes for a single sector is estimated to be approaching 1000 hours.

The time taken to replace just the blankets is the same as for the blankets and divertor because the divertor must be removed first.

The time taken to replace just the divertor cassettes is about 70% of the time for both the divertor and blankets because the port preparation and port sealing operations for the upper and lower divertor port occur in parallel. See figure 2.

Both upper port operations take longer than the lower divertor port operations that occur in parallel, so the divertor replacement is not on the critical path.

The unplanned maintenance time to replace a single blanket varies slightly depending on how many other blankets must be removed first but it is about 1 month on average. The time taken to replace a single divertor cassette is slightly less.

The time to replace the plasma facing components with a single remote handling system is about 22 months. Doubling the number of systems operating in parallel almost halves the maintenance duration so that four systems could complete the maintenance in about 6 months.

A summary of the results from the estimate are shown in table 1 below for 1, 2, 4, 6 and 8 identical remote handling systems operating in parallel to replace all the blankets and divertor cassettes. These figures have been rounded to two significant figures so as not to imply an inappropriate level of accuracy.

| Number of RH systems | Number of sectors per system | Ideal maintenance duration (hours) | Trans-porter delay (hours) | Recovery delay (hours) | 2 shift productivity ratio | Average operator productivity factor | Total elapsed time (hours) | Total elapsed time (months) |
|---|---|---|---|---|---|---|---|---|
| 1 | 16 | 6700 | 0 | 1300 | 0.67 | 0.77 | 15,000 | 22 |
| 2 | 8 | 3300 | 11 | 870 | 0.67 | 0.77 | 8200 | 11 |
| 4 | 4 | 1700 | 15 | 520 | 0.67 | 0.77 | 4300 | 5.9 |
| 6 | 3 | 1200 | 25 | 350 | 0.67 | 0.77 | 3100 | 4.4 |
| 8 | 2 | 800 | 46 | 300 | 0.67 | 0.77 | 2300 | 3.2 |

Table 1. Summary of the maintenance duration to replace the plasma facing components.

The delays due to remote handling equipment failure are approximately 20% of the total, but rising with increasing parallel working. This figure is highly dependent on the reliability figures assumed.

This gives an intrinsic availability of 80% which is high for such a complex system but is due to: redundant or spare systems being provided wherever possible, the relatively high assumed reliability for the highly developed and tested system and because it is assumed that replacement casks are available on demand from the Active Maintenance Facility to replace failed equipment thereby allowing rapid recovery and return to operations.

## 5. Conclusions

### 5.1 Accuracy of the estimate

It is important to consider the accuracy of the data when drawing conclusions from the estimate. The pre-conceptual design does not contain the details required for a more detailed estimate and many of the operations are still unknown.

The estimate should therefore be taken as a first indication only and is best used to show the likely main contributors to the downtime, to allow sensitivity analysis to be conducted and to propose an operating strategy.

It is also important to note that the developing component designs will put pressure on increasing the number and complexity of remote operations and there will be a continuous process to balance these additional demands on the remote handling system against additional maintenance duration, especially during the early stages of the design.

This is where the tool used to estimate the duration will be particularly useful because it is modular in construction and each duration or speed entered in a single location allowing it to be easily modified to analyse the effects of alternative component designs or maintenance strategies.

### 5.2 Optimum duration of planned maintenance

The optimum duration of the planned maintenance is determined by balancing the cost of a faster remote handling system with the cost of reduced power plant availability for generating revenue.

Early estimates of the economics of a fusion power plant of this design were carried out as part of the Power Plant Conceptual Study in 2005. The report [2] suggested a minimum availability of 75% was required.

The DEMO Operational Concept Description for 2012 [7] assumed that the divertor would require replacement every 2 years and the blanket would be replaced every other time the divertor was replaced.

Assuming a 1 month cooling period before remote handing operations begin and a 1 month period to condition and pump-down before start-up, this estimate suggests that four remote handling systems would be required to operate in parallel to meet the 75% availability target.

This maintenance scenario is shown in figure 3 below, in which two 24 month periods of operation occur in a 62 month cycle giving 77% availability.

This does not take into account unplanned maintenance or any other planned maintenance that might arise that cannot be undertaken in parallel with the replacement of the plasma facing components.

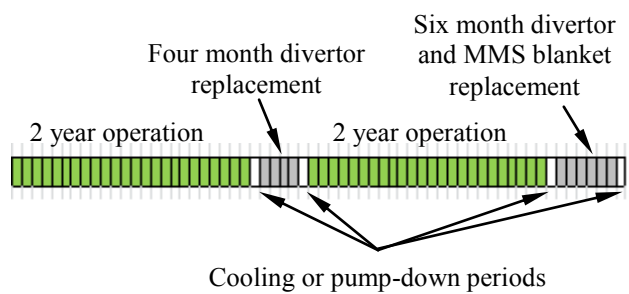

Figure 3: Maintenance cycle to achieve 75% availability for planned maintenance

### 5.3 Further work

The estimation tool must be developed at each stage of the DEMO design process to provide the best data possible to allow the optimum balance to be achieved between the component cost and performance on the one hand and the remote handing and downtime costs on the other.

Improved input data is required to allow better conclusions to be drawn from the output, particularly in terms of operation duration and failure rates.

During 2013 the model will be input into a logistics software package by the KIT Institute for Conveying Technology and Logistics, to allow additional refinement and analysis to be carried out.

## Acknowledgments

This work is funded by the RCUK Energy Programme [grant number EP/I501045] and by EFDA. To obtain further information on the data underlying this paper please contact *PublicationsManager@ccfe.ac.uk*. The views and opinions expressed herein do not necessarily reflect those of the European Commission.

## References

[1]   M. Coleman, N. Sykes, D. Cooper, D. Iglesias, R. Bastow, A. Loving, J. Harman, *Concept for vertical maintenance remote handling system for multi-module blanket segments in DEMO*, published in the proceedings of ISNFT-11, 2013.

[2]   D. Maisonnier, I Cook, P Sardain, R. Andreani, L. Di Pace et al, *A conceptual study of commercial fusion power plants*, EFDA_D_2JP3X3, 2005.

[3]   B. Conway, *Remote Handling Maintenance Compatibility Assessment of the HTS and LTS Multi-Module Concepts,* EFDA_D_2L9GW5, 2012.

[4]   J. Thomas, A. Loving, O. Crofts, R. Morgan, J. Harmon, *DEMO active maintenance facility concept progress 2012*, published in the proceedings of ISNFT-11, 2013.

[5]   O. Crofts, *Duration of Remote Maintenance Operations for the DEMO VMS,* EFDA_D_2LMX3M, 2012.

[6]   S. Collins, G. Matthews, J. Thomas, G. Herman, *Factors affecting remote handling productivity during installation of the ITER-like wall at JET*, published in the proceedings of 27th SOFT, 2012.

[7]   A. Loving, P. Allan, N. Sykes, P. Murcutt, *Development and application of high volume remote handling systems in support of JET and ITER*, published in the proceedings of ISFNT-10, 2011.

[8]   J. Harman, *WP12 DEMO Operational Concept Description*, EFDA_D_2LCY7A, 2012.